\documentclass{article}
\usepackage{emulateapj,graphics,onecolfloat}
\received{2001 November 26}
\begin{document}

\twocolumn
[
\title{Completeness in Weak Lensing Searches for Clusters}

\author{Martin White\altaffilmark{1}, Ludovic van Waerbeke\altaffilmark{2,3},
Jonathan Mackey\altaffilmark{4}}

\altaffiltext{1}{Departments of Physics and Astronomy, University of California,
Berkeley, CA 94720}

\altaffiltext{2}{Institut d'Astrophysique de Paris, 98bis boulevard Arago, 75014
Paris France}

\altaffiltext{3}{Canadian Institut for Theoretical Astrophysics, 60 St Georges St,
Toronto, M5S 3H8 Ontario, Canada}

\altaffiltext{4}{Harvard-Smithsonian CfA, 60 Garden St, Cambridge, MA 02138}

\affil{email: mwhite@astron.berkeley.edu}

\begin{abstract}
Using mock observations of numerical simulations, we investigate the
completeness and efficiency of searches for galaxy clusters in weak
lensing surveys.  While it is possible to search for high mass objects
directly as density enhancements using weak lensing, we find that
line-of-sight projection effects can be quite serious.
For the search methods that we consider, to obtain high completeness
requires acceptance of a very high false-positive rate.  The false positive
rate can be reduced only by significantly degrading the completeness.
Both completeness and efficiency are dependent upon the filter used in the
search and the desired mass threshold, emphasizing that a measurement of the
3D mass function from gravitational lensing is affected by a number of
biases which mix `cosmological' and observational issues.
\end{abstract}

]

\section{Introduction}

Clusters of galaxies are one of our most important cosmological probes.
As the most recent objects to form in the universe their number density and
properties are exquisitely sensitive to our modeling assumptions.
Their composition accurately reflects the mix of matter in the universe.
They are bright and can be ``easily'' seen to large distances, allowing
constraints on the crucial interval $0< z\la 1$ where the universal
expansion changes from deceleration to acceleration.
They are located close to their formation site.
Being bright and sparse they are excellent tracers of the large-scale
structure -- they are highly biased so their clustering is easy to measure
and is much more straightforwardly computed from theory than that of
galaxies.

For many years it has been realized that a large, homogeneous, sample of
clusters would impose particularly strong constraints on our model of
structure formation, provided all clusters above a given mass threshold
were included.
In a recent manifestation of this idea
Haiman, Mohr \& Holder~(\cite{HaiMohHol})
have suggested using the counts of clusters of galaxies above a certain mass
threshold to probe the evolution of the dark energy believed to be causing
the acceleration of the universal expansion.
A more pedestrian goal would be to constrain the matter density and amplitude
of the mass fluctuations in the universe
(e.g.~Holder, Haiman \& Mohr~\cite{HolHaiMoh} for a recent discussion).
A study of the clustering of such samples could provide another measurement
of the angular diameter distance-redshift relation
(Cooray et al.~\cite{CooHuHutJof})
or the linear growth rate of fluctuations.

However, constructing large samples of galaxy clusters for statistical
analyses remains a difficult task.  For many years the only available
catalogues were based on projected galaxy overdensity, though it was
quickly realized that such samples suffer from projection effects and
the large scatter between optical richness and cluster mass
(for recent theoretical studies see
e.g.~van Haarlem, Frenk \& White \cite{vHaFreWhi};
     Reblinsky \& Bartelmann \cite{RebBar}
     White \& Kochanek \cite{WhiKoc}).
At present cluster samples have been constructed using optical, X-ray and
Sunyaev-Zel'dovich `surveys' each with their own advantages and
disadvantages.

In the last few years it has become possible to search for high mass
objects directly as density enhancements using weak gravitational lensing.
Since the first detections by Bonnet, Mellier \& Fort (\cite{BonMelFor})
and Fahlman et al.~(\cite{FKSW}), this technique has been demonstrated
many times on previously known clusters (see Wu et al.~\cite{WCFX} and
Hattori et al.~\cite{HKM} for recent reviews).
Erben et al.~(\cite{erben00}) and Umetsu \& Futamase (\cite{UmeFut}) have
reported `dark clumps', where mass concentrations are seen with no
optical or X-ray detection of a cluster.
To date there is only one case of a confirmed cluster discovered first with
weak gravitational lensing:
Wittman et al.~(\cite{Witetal01}) serendipitously detected a $z=0.276$ cluster
with $\sigma_v\sim 600$km/s in the corner of one of the fields they used
previously for cosmic shear.

Thus lensing offers a completely new way to find clusters.
It has been claimed by some authors that weak lensing offers our first chance
to construct a truly mass selected sample of clusters.
As with all new methods, lensing avoids some of the problems which plague
other methods but has its own systematics.
In this paper we want to look at the power and pitfalls of this method.
Specifically we want to look at how efficient and complete weak lensing
surveys are at constructing a mass selected sample of clusters, assuming
ideal observing conditions.

\begin{figure*}
\begin{center}
\resizebox{7in}{!}{
\includegraphics{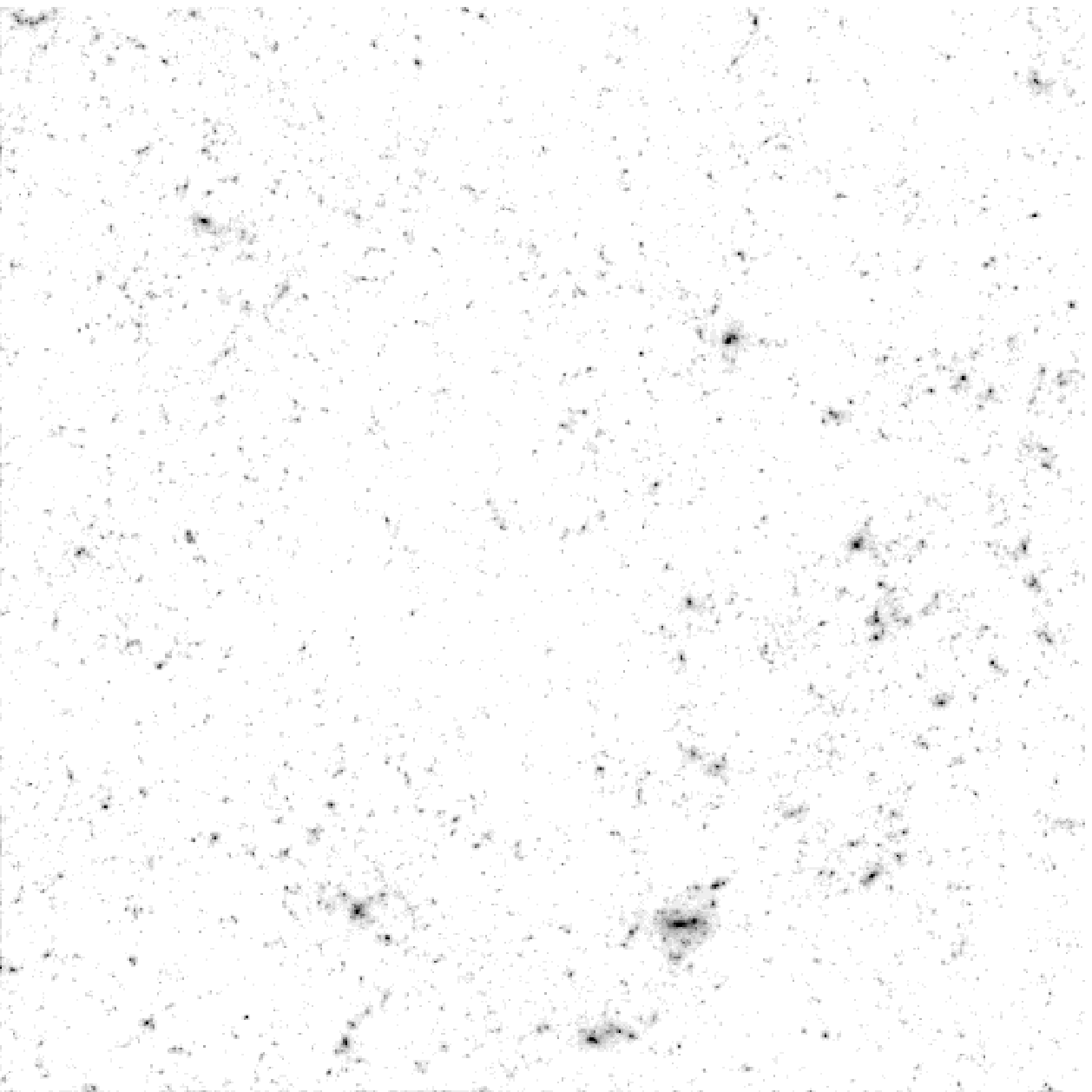}
\includegraphics{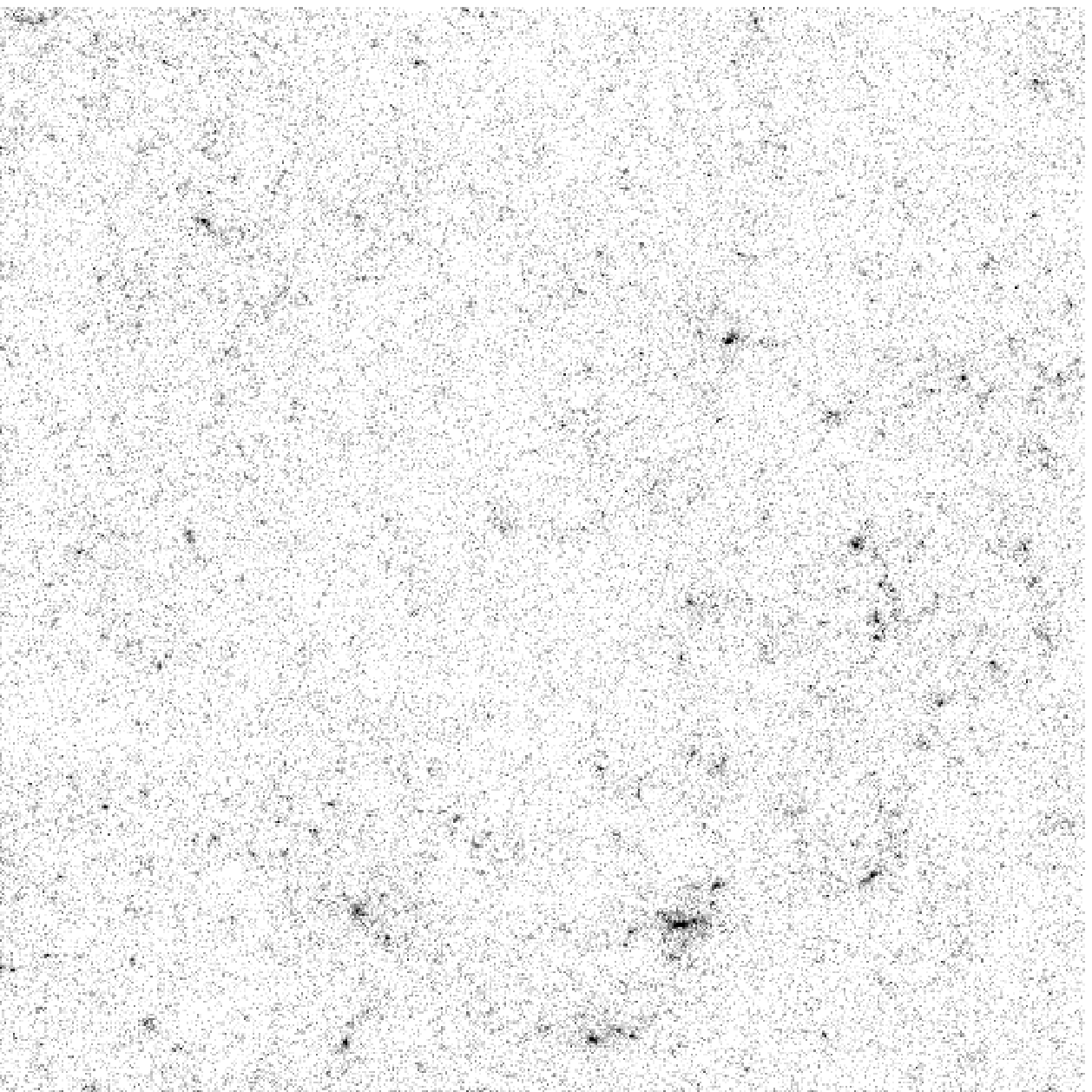}}
\resizebox{7in}{!}{
\includegraphics{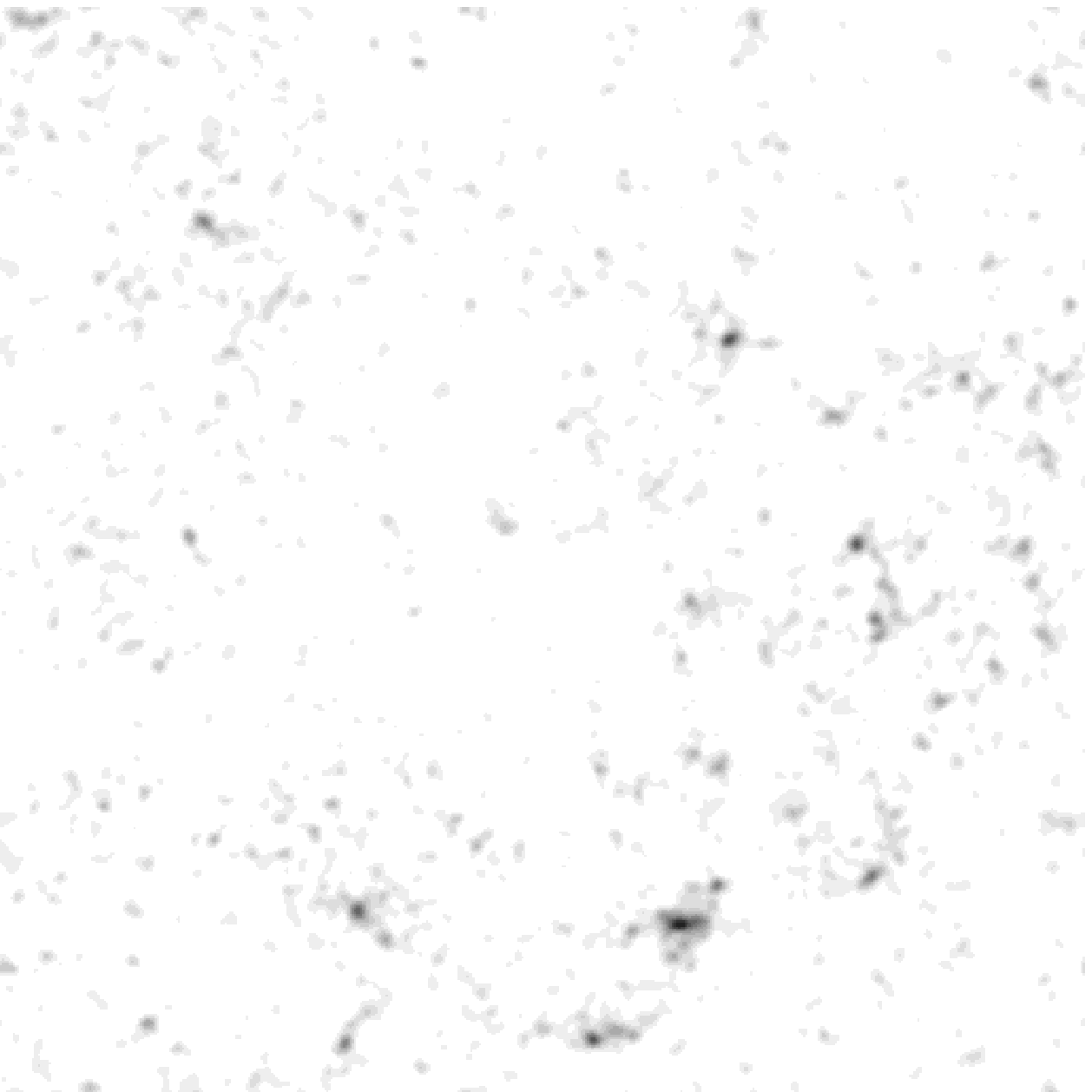}
\includegraphics{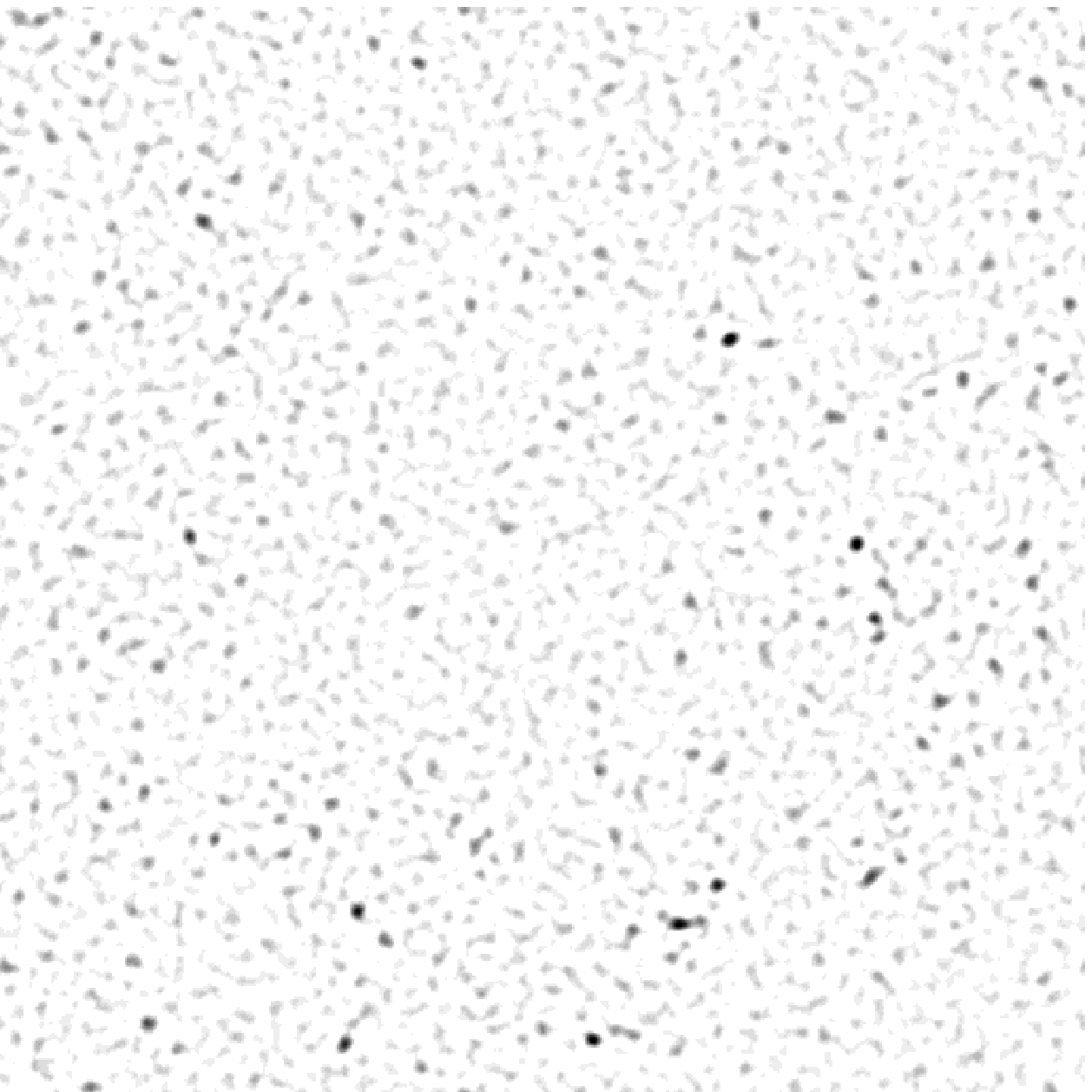}}
\end{center}
\caption{\footnotesize%
(top left) A simulated $\kappa$ map, $5^\circ$ on a side, with a linear
greyscale which maps all pixels with $\kappa<0$ to white.
(top right) The same field with noise added.
(bottom left) The noisy map, smoothed with a Gaussian of $1'$.
(bottom right) The $M_{\rm ap}$ map obtained from the noisy map with
a filter of about $5'$.}
\label{fig:fourmaps}
\end{figure*}

\section{Simulated observations}

\subsection{N-body simulation} \label{sec:nbody}

We wish to make simulated convergence maps which are as realistic as possible
while at the same time knowing the real 3D location of any clusters in the
field.
Our programme begins with a model for the evolution of the dark matter
which governs the formation of large-scale structure.  On Mpc scales we expect
that the baryonic matter will faithfully trace the dark matter, thus our model
should reproduce the spatial distribution of mass.
This problem can be well tackled by modern numerical simulations.
We have run a $512^3$ particle simulation of a $\Lambda$CDM model in a
$300h^{-1}$Mpc box using the {\sl TreePM-SPH\/} code
(White et al.~\cite{TreePM}) operating in collisionless (dark matter only)
mode.  This simulation represents a large cosmological volume, to include
a fair sample of rich clusters, while maintaining enough mass resolution
to identify galactic mass halos.
Because it provides a reasonable fit to a wide range of observations,
including the present day abundance of rich clusters of galaxies
(Pierpaoli, Scott \& White \cite{PieScoWhi}), we have simulated a
$\Lambda$CDM cosmology with $\Omega_{\rm m}=0.3$,
$\Omega_\Lambda=0.7$, $H_0=100\,h\,{\rm km}{\rm s}^{-1}{\rm Mpc}^{-1}$
with $h=0.7$, $\Omega_{\rm B}=0.04$, $n=1$ and $\sigma_8=1$.
The simulation was started at $z=60$ and evolved to the present with the
full phase space distribution dumped every $100h^{-1}$Mpc from $z\simeq 2$ to
$z=0$.
The gravitational force softening was of a spline form, with a
``Plummer-equivalent'' softening length of $20\,h^{-1}$kpc comoving.
The particle mass is $1.7\times 10^{10}h^{-1}M_\odot$ allowing us to find
bound halos with masses several times $10^{11}h^{-1}M_\odot$ and giving
several tens of thousands of particles in a cluster mass halo
($>10^{14}h^{-1}M_\odot$) to begin to resolve substructure.

For every output of the simulation we produce a halo catalogue by running a
``friends-of-friends'' group finder (e.g.~Davis et al.~\cite{DEFW}) with a
linking length $b=0.15$ (in units of the mean interparticle spacing).
This procedure partitions the particles into equivalence classes, by linking
together all particle pairs separated by less than a distance $b$.
We use $0.15$ rather than the more canonical $b=0.2$ as we find that the larger
linking length frequently merges what we would by eye list as separate halos.
By effectively `removing' one of the halos from our list this would have a
deleterious impact upon our efficiency calculations.
While this effect is not entirely eliminated when using $b=0.15$, it is
significantly reduced.
We keep all groups above 64 particles, which imposes a minimum halo mass of
order $10^{12}h^{-1}M_\odot$.
For each identified halo we compute, using the 3D distribution of all of the
particles in the simulation, the mass (we use $M_{200}$, the mass enclosed
within a radius, $r_{200}$, within which the mean density is 200 times the
{\it critical\/} density at that redshift), velocity dispersion etc.~so we
can understand our selection in terms of the intrinsic, rather than projected,
cluster properties.

\begin{figure}
\begin{center}
\resizebox{3.5in}{!}{\includegraphics{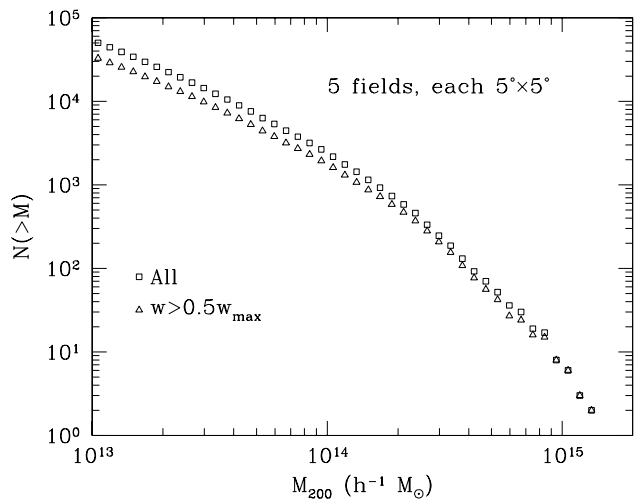}}
\end{center}
\caption{\footnotesize%
The `projected' mass function of clusters in our five $5^\circ\times 5^\circ$
simulated lensing fields.  The squares indicate all clusters lying between
the source and the observer in the fields, the triangles those for which
the lensing kernel, $t(1-t)$, is greater than half its peak value.
We expect that extreme projection effects will be smaller for this restricted
sample of clusters, since their weight in the lensing map is larger.}
\label{fig:massfn2d}
\end{figure}

\subsection{The convergence map}

To make a weak lensing map we need to perform an integral of the mass density
(times a weight function) back along the line-of-sight to the redshift of the
furthest source $z_{s,{\rm max}}$.
Since the sources are at a cosmological distance, while our simulation volume
is relatively small, we do this by ``stacking'' different slices through
the box at earlier and earlier output times (see Fig.~1 of da Silva
et al.~\cite{dSBLT} for a diagram of this approach in another context).
Specifically we divide the simulation box at every output up into three pieces
of $300\times 300\times 100h^{-1}$Mpc by trisecting in the line-of-sight
dimension.
A given observational field is then obtained by stepping back from $z=0$ to
$z=z_{s,{\rm max}}$ choosing, every $100h^{-1}$Mpc along the line-of-sight,
the density field from one third of the box (see below) at that output.
To avoid repeatedly tracing through the same structure we shift the simulation
volme perpendicular to the line-of-sight by a random amount and wrap using the
periodicity of the simulation volume.  All of the mass in that third of the box
is projected onto the sky in this manner.
The convergence is
\begin{equation}
  \kappa = {3\over 2}\Omega_{\rm mat}(H_0D_*)^2 \int dt\ w(t)\ {\delta\over a}
\end{equation}
where $D_*$ is the (comoving) angular diameter distance,
\begin{equation}
  {dD\over da}\equiv {1\over a^2H(a)} \qquad ,
\end{equation}
to a fiducial point beyond the furthest source,
$t=D/D_*$ is a dimensionless distance, $w(t)$ is a weight function
\begin{equation}
  w(t) = \int_t^1 dt_s\, {dn\over dt_s}\ {t(t_s-t)\over t_s}
\end{equation}
which reduces to $t(1-t)$ if all of the sources are at a fixed distance $D_*$
and $\delta=\rho/\bar{\rho}-1$.
We approximate this integral as a weighted sum of the projected mass $\Sigma$
in each plane, with the projection being done parallel to the edges of the
simulation volume.
Studies by Hamana, Colombi \& Mellier (\cite{HamColMel}) have indicated that
this method produces results which are almost identical to more numerically
intensive ray tracing simulations.
We have found that there is a slight positive bias in the power spectrum of
$\kappa$ using this method as compared to the ``tiling'' method of
White \& Hu (\cite{WhiHu}) which does not approximate the path with segments
parallel to the box boundaries.
This small bias will not affect any of our conclusions.

We have chosen $100h^{-1}$Mpc as our sampling interval because it is large
enough that edge effects are minimal even for rich clusters while being fine
enough that line-of-sight integrals are well approximated by sums over the
(static) outputs.  However, even though only a small fraction of clusters lie
within $r_{200}$ of a slice boundary, we decided to require that the
orientation and offset change only on every third slice.  Thus if we choose
at one redshift the front of the box the next slice is required to be the
middle and the next the back.
In this manner the structure is continuous across $300h^{-1}$Mpc distances,
but still evolves in steps of $100h^{-1}$Mpc not $300h^{-1}$Mpc.

As a first step we have generated 5 maps, each $5^\circ\times 5^\circ$
and $1024^2$ pixels, with sources fixed at $z_s=1$ (one example is shown
in Fig.~\ref{fig:fourmaps}).
A survey which is significantly shallower suffers a loss of lensing signal and
becomes more sensitive to systematics related to intrinsic alignments of
galaxies.  A much deeper survey is extremely expensive of telescope time.
Thus $z_s\sim 1$ has been chosen by most workers in the field.
By choosing all of our sources to lie at precisely $z_s=1$ we obtain a good
first approximation to planned or existing surveys, while at the same time
decoupling any uncertainties in the source distribution from the effects we
are concerned with in this work.
Our maps are only approximately independent as they are drawn from the same
simulation, but the random orientations allow us to sample different possible
projection effects.
For each map we can add Gaussian pixel noise at a specified level and we
smooth the maps (using fourier transform methods) with a Gaussian beam of
a specified FWHM or convolve the map with some other filter.

\subsection{The group catalog}

\begin{figure}
\begin{center}
\resizebox{3.5in}{!}{\includegraphics{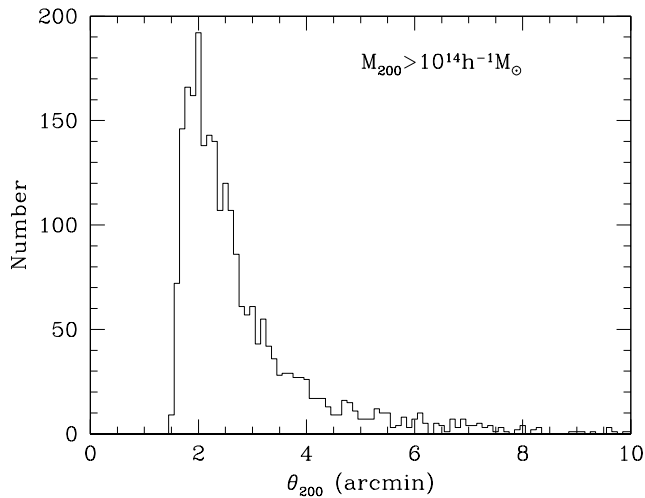}}
\end{center}
\caption{\footnotesize%
The distribution of projected `virial' radii, $\theta_{200}$, for the clusters
in our five $5^\circ\times 5^\circ$ simulated lensing fields with
$M_{200}>10^{14}h^{-1}M_\odot$.  We define $\theta_{200}$ as the angle
subtended by $r_{200}$ at the distance of the cluster.}
\label{fig:clussize}
\end{figure}

The same random slices and offsets are used to project the group catalogue
into the field, and to produce 3D images\footnote{The particle distribution is
visualized using {\sl Tipsy\/} from the N-body shop at the University of
Washington.} of the field of view from the particle distribution
(which we use to check for projection effects).
We shall denote by `cluster' any halo having $M_{200}>10^{14}h^{-1}M_\odot$
and our catalog includes all such halos lying in the field with $z\le 1$.
We define the center of a cluster as the position of the potential minimum,
calculating the potential using only the particles in the FoF group.  This
proved to be more robust than using the center of mass, as the potential
minimum coincided closely with the density maximum for all but the most
disturbed clusters.
The mass function of clusters lying between the source and observer in our 5
simulated fields is shown in Fig.~\ref{fig:massfn2d}.  The distribution of
angular sizes is shown in Fig.~\ref{fig:clussize}.

\subsection{Finding \& matching peaks}

For a given (possibly noisy and smoothed) map we need an algorithm for
finding peaks.  We have chosen a simple strategy whereby we search the map
for all pixels which are a local maximum (this is essentially the procedure
used on real data).  This set forms our base peak list, which we number.
We then search around each maximum and include the adjacent pixels if
$\kappa > f \kappa_{\rm max}$ where $f$ is a user specified fraction
typically set to 0.7.  The peaks are all extended at the same rate, so that
adjacent peaks do not swallow each other.
This algorithm then returns for every pixel in the map the peak number
(or possibly ``no peak'') to which it belongs.
For each peak we keep track of both the maximum $\kappa$ and the sum,
$\kappa_{\rm sum}$, of the convergence in all pixels above the threshold
$f\kappa_{\rm max}$.  The latter is loosely correlated with mass.
The exact extent of the peaks and the precise definition of $\kappa_{\rm sum}$
will not affect our results.

In addition to our `peak list' we have our `cluster list' from the 3D
cluster catalogues.  Given the very large number of both peaks and clusters,
and the lack of distance information in the peak list, matching these can be
quite problematic.
We perform the match in two directions: whether a peak is in the cluster list
(forward match) and whether a cluster is in our peak list (backward match).
We typically find that all clusters above $10^{14}h^{-1}M_\odot$ lie within
$\pm 1$ pixel of an extended peak, but some clusters lie near peaks (local
maxima) with very low values of $\kappa$ (see below).
Because each of the lists is so long and we are only using 2D information in
associating peaks to clusters, a `match' is claimed only if the forward and
backward associations agree.

\begin{figure}
\begin{center}
\resizebox{3.5in}{!}{\includegraphics{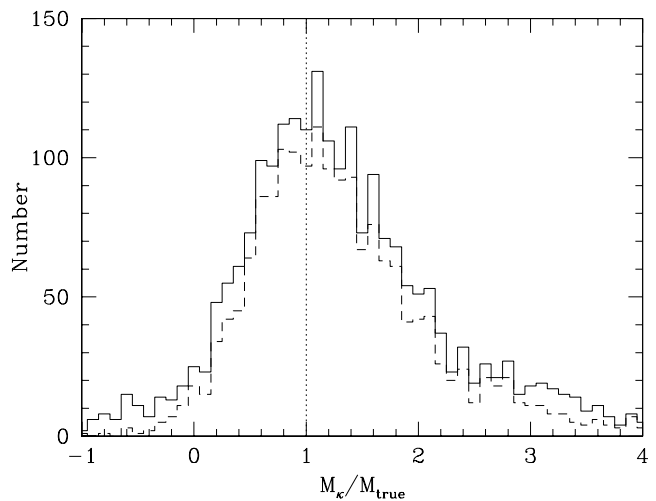}}
\end{center}
\caption{\footnotesize%
The ratio of the ``lensing mass'' to the true mass for clusters above
$10^{14}h^{-1}M_\odot$ for our five $5^\circ\times 5^\circ$ lensing fields.
The solid histogram is all clusters and the dashed line is those for which
$w>w_{\rm max}/2$ (see text).}
\label{fig:mhist}
\end{figure}

The code produces a list of peaks and halos with matches flagged (plus cases
with only a forward or backward match).  There are two key numbers which we
shall focus on below.
The first is the fraction of peaks which matched at least one halo, which
will determine our `efficiency'.
The second is the fraction of halos which matched at least one peak, which
will determine our `completeness'.

Note that as we begin to smooth the maps and add noise, the 1-1 correspondence
between peaks and halos will begin to degrade.  We take the attitude that
all potential detections would be followed up with e.g.~X-ray observations
or redshifts.  Thus if two halos match a single extended peak we can count
those both as detections since followup of that area of sky would presumably
find both of them.

Our method is very robust and can be easily automated, but it is not perfect.
In particular it can cause us to underestimate the efficiency of a lensing
search due to the way we define our halos.  For example, if two massive halos
are close together (perhaps in the initial stages of a merger or interaction)
they can be linked by our group finder into a single halo.  Our group catalog
will then contain parameters only for the mass around the potential minimum,
missing the other halo entirely.  This halo still has a large amount of mass
associated with it however, and is quite overdense, so it will likely
correspond to a $\kappa$ peak.  This peak will be erroneously counted as a
miss as there is no corresponding entry in the cluster list.

This effect is mitigated to a large extent by the relatively small linking
length we have chosen to define our 3D catalog.  Neighboring halos are linked
only if the material between them is $\ga 10^2\times$ overdense.
We have found one example of such an artificial linking for systems of
cluster mass, corresponding to a chain of halos lying along a filament.
This single system changes our results by less than a percent.
Inspection of other `strange' peaks has not yielded any other examples of
this effect.

\begin{figure*}
\begin{center}
\resizebox{7in}{!}{
\includegraphics{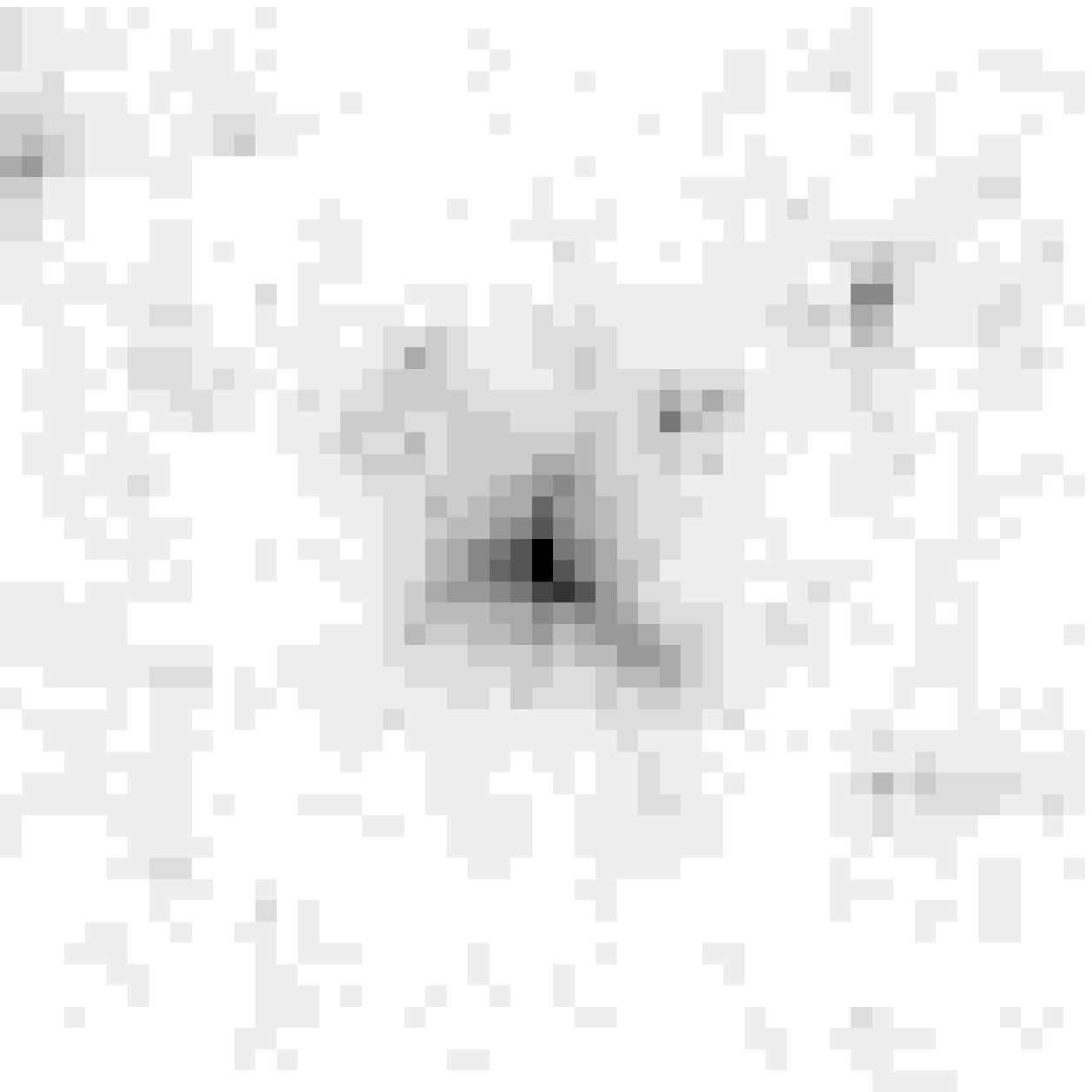}
\includegraphics{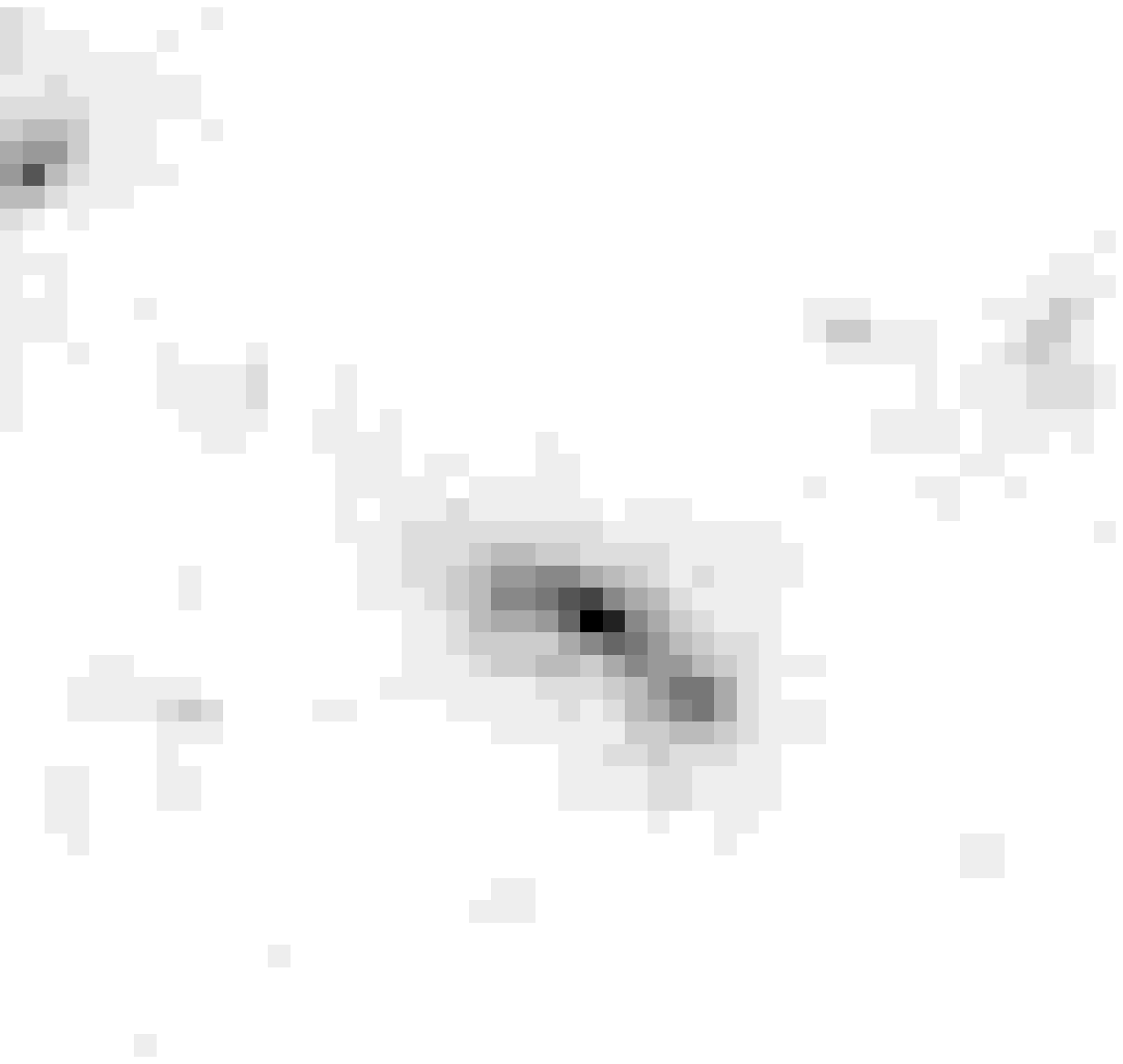}
\includegraphics{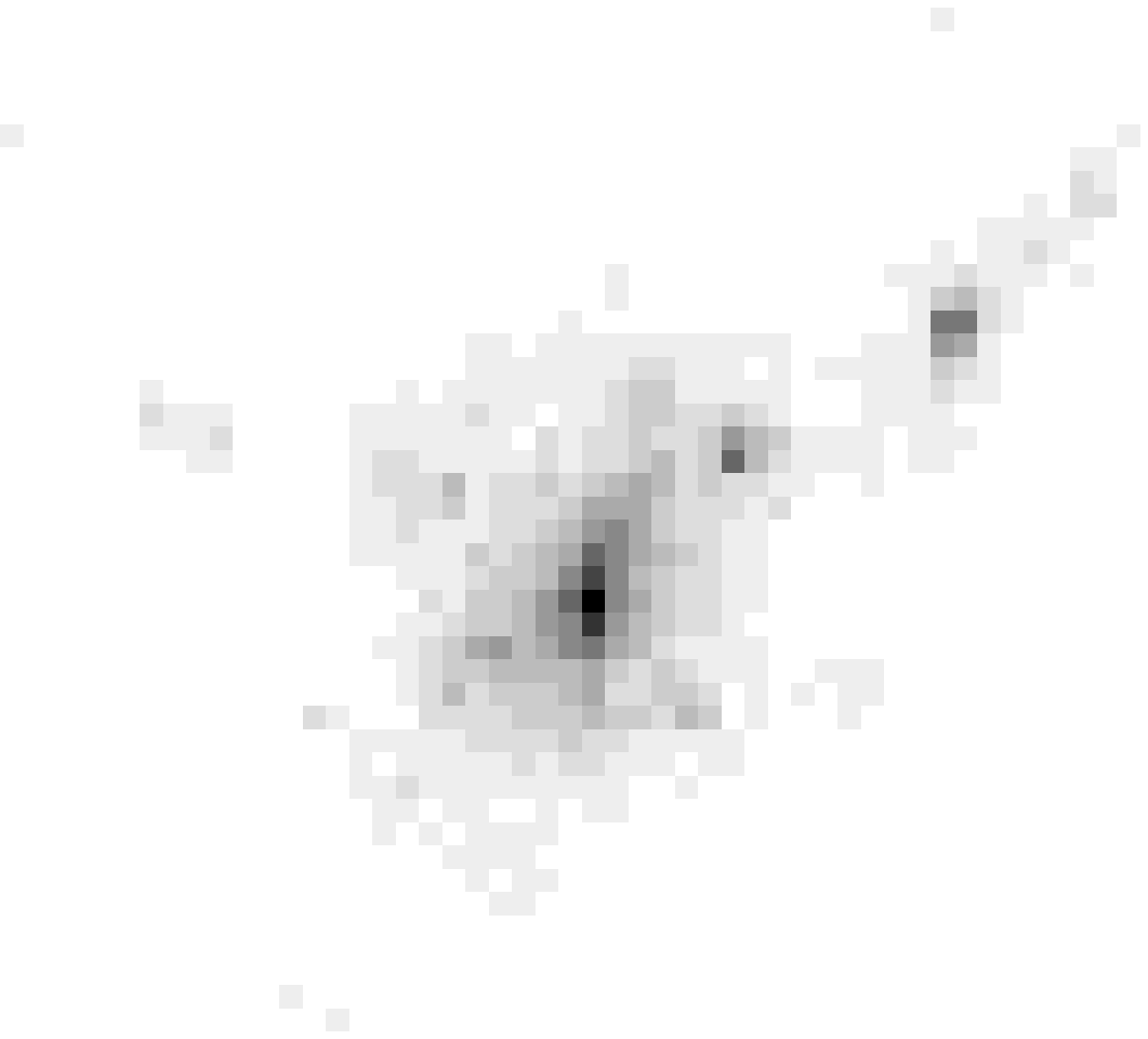}}
\end{center}
\caption{\footnotesize%
A zoom in of one of our convergence maps showing two clusters which lie
almost on top of each other in projection.
(left) the full $\kappa$ map.
(middle) the portion coming from the $100h^{-1}$Mpc slice at low redshift.
(right) the portion coming from the $100h^{-1}$Mpc slice at higher redshift.
The regions shown are $0.3^\circ$ on a side.}
\label{fig:overlap}
\end{figure*}

\section{Results}

\subsection{Raw maps}

\begin{table}
\begin{center}
\begin{tabular}{c|c|cc}
 & Peaks & \multicolumn{2}{c}{Clusters} \\
 &         & $w>0$  & $w>w_{\rm max}/2$ \\ \hline
 All       & 402154 & 0.99 & 0.99 \\
$>0.0$     & 288933 & 0.98 & 0.99 \\
$>0.2$     &   4712 & 0.62 & 0.81 \\
$>0.4$     &    629 & 0.21 & 0.28
\end{tabular}
\end{center}
\caption{\footnotesize%
Peak statistics for the `raw' maps, as a function of threshold
$\kappa_{\rm max}$.
The columns give the number of peaks above the thresholds listed, and the
fraction of clusters (halos with $M_{200}>10^{14}h^{-1}M_\odot$) satisfying
the lensing kernel cut which were found by matching to those peaks.
There were 2415 and 1775 clusters with $w>0$ and $w>w_{\rm max}/2$
respectively.}
\label{tab:raw}
\end{table}

We first present results on the raw $\kappa$ maps, i.e.~without adding noise
or smoothing the map.  These results will set a theoretical `best' performance
and allow us to understand how much things are degraded by noise and smoothing.
Our results for the noise-free, unsmoothed maps are listed in
Table~\ref{tab:raw}.
We see that there are a very large number of peaks, and that almost all
clusters can be matched to peaks (we discuss the rare misses below).
If we apply a threshold to our peak finder to reduce the peaks to a
reasonable number our completeness begins to drop rapidly.

The reason for this can be seen by considering the distribution of $\kappa$
values in the region of known clusters.  We have chosen to do this in the
context of cluster mass measurements, realizing that this issue has been
discussed in many contexts previously.
As a first step we went through our cluster list and for all clusters above
$10^{14}h^{-1}M_\odot$ we summed the values of $\kappa$ within a disk centered
on the known cluster center and with radius equal to the known value of
$r_{200}$.  Using the cluster redshift we then converted this to a mass.
A histogram of the mass compared to the true mass is shown in
Fig.~\ref{fig:mhist} where we see that there is both a large scatter, as has
been emphasized before\footnote{See for example the discussion in
\S7.3 of Metzler, White \& Loken (\cite{MetWhiLok}).}, and a positive offset.
Perhaps the most dramatic are the few clusters with a negative mass.  These
arise when a cluster in a region where the lensing kernel is small has a
large void projected along the line-of-sight at a distance where the lensing
kernel is large.  This leads to a negative mean $\kappa$ even though there
exists a significant mass overdensity.
To isolate this effect we further require that the lensing kernel at the
cluster position be more than half of its peak value.  This largely removes
the negative tail, although even with this cut 34 clusters (out of more than
1750) with $M_{\kappa}<0$ remain.

The tail to very large mass ratios occurs when a massive cluster is projected
on top of a less massive cluster.  These lines-of-sight were excluded in the
analysis of Metzler et al.~(\cite{MetWhiLok}), but we have not done so here.
Thus our procedure will return something close to the cumulative mass for each
of these systems, which will be a large overestimate for the low mass system.
We show an example of such an overlap in Fig.~\ref{fig:overlap}.
Again, restricting ourselves to clusters near the peak of the lensing kernel
reduces this effect.
We shall return to the projection effects on the mass measurement
of the detected peaks at the end of \S\ref{sec:S-statistic}.

While the details change, this kind of effect is also seen in the distribution
of peak $\kappa$ values or in other measures of `mass' based on lensing.

\subsection{Adding noise}

While the above results can serve to indicate some of the pitfalls in weak
lensing searches for clusters, they do not indicate how a realistic search
would perform.  To make further steps in this direction we need to include
`noise' in our maps.
We will consider an `ideal' or optimistic survey for the purposes of
highlighting the cosmological, rather than observational, effects.
Thus we shall neglect sources of noise intrinsic to the telescope or detector
system and focus instead on the irreducible `noise' due to galaxy properties.
Following van Waerbeke (\cite{vWae}) we model the noise in a $\kappa$ map
that would arise from processing a shear map using a technique such as that
of Kaiser \& Squires (\cite{KaiSqu}) as Gaussian, correlated only by any
smoothing kernel applied to the map.  If the mean intrinsic ellipticity of
the source galaxies is $\gamma_{\rm int}$ then the noise introduced in our
$\kappa$ map has variance
\begin{equation}
  \sigma_{\rm pix}^2 = {\gamma_{\rm int}^2\over \bar{n}\theta_{\rm pix}^2}
\end{equation}
where $\bar{n}$ is the mean number density of sources.
Assuming $\gamma_{\rm int}=0.2$ (which is slightly optimistic for ground
based work;
Kaiser~\cite{Kai98}; Crittenden, Natarajan, Pen \& Theuns~\cite{CNPT}) and
$\bar{n}=25$ gal/arcmin${}^2$ (again slightly optimistic) we have
$\sigma_{\rm pix}\simeq 0.14$ for our $0.3'$ pixels.
This level of noise is quite optimistic, compared to current observations,
which will make our conclusions conservative.  None of our main results will
depend on our specific choice of $\sigma_{\rm pix}$, though the absolute
meaning of our signal-to-noise cuts will clearly scale with
$\sigma_{\rm pix}$.

Even this level of noise is quite large compared to our signal, so we need
to smooth the maps to enhance the contrast of our signal to the noise.
We have not attempted to search for an `optimal' filter, matched to the
predicted shape of the cluster.  Instead we have chosen to either smooth
the maps with a Gaussian whose FWHM is roughly matched to the size
of clusters at cosmological distances, $1'-2'$, or to apply the $M_{\rm ap}$
filter already discussed in the literature (see \S\ref{sec:S-statistic}).

\begin{figure}
\begin{center}
\resizebox{3.5in}{!}{\includegraphics{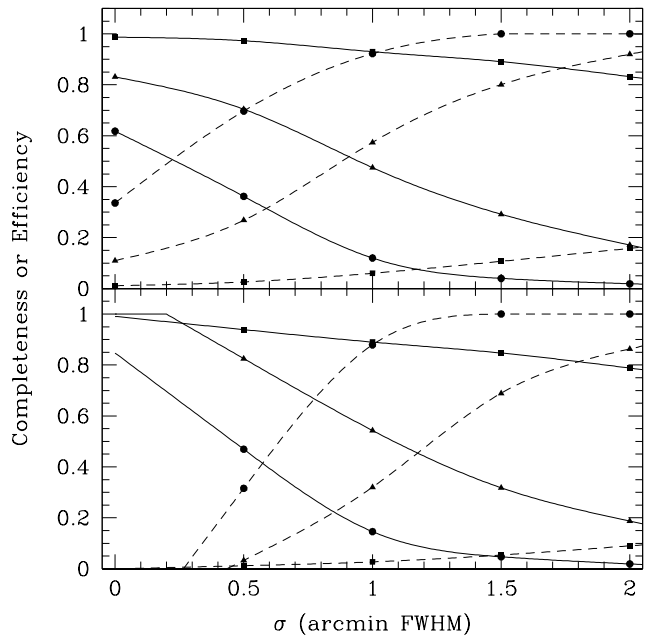}}
\end{center}
\caption{\footnotesize%
Completeness and efficiency as a function of smoothing for maps with
(top) no noise and (bottom) noise of rms 0.14 per pixel.
Solid lines are the fraction of clusters (groups with
$M_{200}>10^{14}h^{-1}M_\odot$) in the 5 fields which are matched
(completeness), dashed lines are the fraction of identified peaks which
correspond to clusters (efficiency).  The symbol types denote the $\kappa$
threshold for counting peaks: (squares) $\kappa_{\rm max}>0$,
(triangles) $\kappa_{\rm max}>0.1$, (circles) $\kappa_{\rm max}>0.2$.}
\label{fig:smth}
\end{figure}

The top panel of Fig.~\ref{fig:smth} shows how a survey's completeness and
efficiency depend on the smoothing scale in the case of no noise.
As expected increasing the smoothing decreases the number of false positives
and thus the expense of follow up observations; but it also decreases the
survey completeness.
The best match of peak threshold and smoothing will depend on the trade offs
between these two issues.
The trade off is also affected by the level of noise in the map, as is shown
in the bottom panel of Fig.~\ref{fig:smth}.
Note that for no combination of parameters is it possible to have $>50\%$
completeness with $<50\%$ contamination!
This may be traced primarily to the large scatter in the mass-peak relation
shown in Fig.~\ref{fig:mhist} (for a discussion of precisely this effect in
redshift surveys for clusters, see White \& Kochanek \cite{WhiKoc}).

We note that only for very high thresholds is our efficiency close to 100\%.
This means that there are quite prominent peaks in the maps which do not match
any cluster with $M_{200}>10^{14}h^{-1}M_\odot$.
This could have relevance to the question of `dark clusters' (see
Fischer \cite{fischer99}, Erben et al.~\cite{erben00},
Umetsu \& Futamase \cite{UmeFut}).

\subsection{Matched filter} \label{sec:S-statistic}

One can enhance the signal-to-noise for cluster-like structures by convolving
the $\kappa$ map with a `matched filter'.  For a certain class of such filters,
known as aperture mass measures
(Schneider \cite{S96}; Schneider et al.~\cite{SvWJK})
this operation is easy to implement directly on the shear field itself:
convolution of the $\kappa$ map with a kernel $U$ can be shown to be the same
as convolving the tangential shear map with a related kernel $Q$ provided
the kernel $U$ vanishes outside of some radius $\vartheta$ and is compensated,
viz
\begin{equation}
  \int \theta d\theta\ U(\theta) = 0 \qquad .
\end{equation}
The convolution of $\kappa$ with $U$ to produce the $M_{\rm ap}$ map is a
bandpass filter on the convergence map, with the filter function quite narrow
in (spatial) frequency space (Bartelmann \& Schneider \cite{BarSch}).
Thus we may expect that for an appropriately chosen $M_{\rm ap}$ scale,
$\vartheta$, we will enhance the contrast of clusters in the map.

We have created, from our noisy $\kappa$ maps, a series of $M_{\rm ap}$ maps
with different filtering scales.  We use the simplest, $\ell=1$, $M_{\rm ap}$
kernel
\begin{equation}
  \theta^2 U(\theta) = {9\over \pi}(1-u^2)\left({1\over 3}-u^2\right)
  \quad {\rm if}\ \theta<\vartheta.
\end{equation}
Here $u\equiv \theta/\vartheta$ and $\vartheta$ is the filter size.
We perform this convolution directly on the $\kappa$ map, with $M_{\rm ap}$
filters for which $\vartheta$ is a multiple of the pixel scale.
A typical cluster is $\sim 10$ pixels across, so pixelization effects are not
too severe on the scales of interest.
Convolution of the $\kappa$ map with this kernel indeed enhances the visibility
of clusters in the maps as can be seen in Fig.~\ref{fig:fourmaps}.
Finally we produce $S-$statistic maps by dividing our $M_{\rm ap}$ maps by
the rms fluctuation of the {\it noise\/} map.

\begin{figure}
\begin{center}
\resizebox{3.5in}{!}{\includegraphics{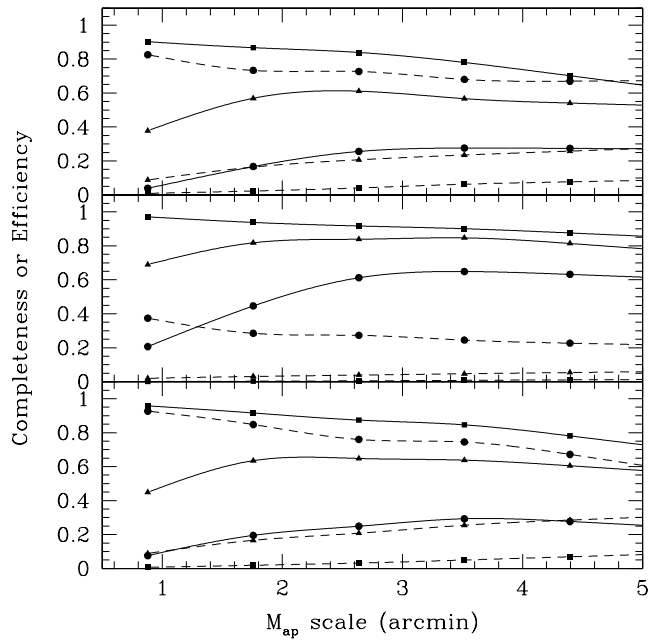}}
\end{center}
\caption{\footnotesize%
Completeness and efficiency as a function of filtering scale for $S-$statistic
maps with noise added as in Fig.~\protect\ref{fig:smth}.
(top) All clusters above $10^{14}h^{-1}M_\odot$
(middle) all clusters above $3\times 10^{14}h^{-1}M_\odot$ and
(bottom) clusters above $10^{14}h^{-1}M_\odot$ with $w>w_{\rm max}/2$.
Solid lines are the fraction of clusters in the 5 fields which are matched
(completeness), dashed lines are the fraction of identified peaks which
correspond to clusters (efficiency).  The symbol types denote the $S$
threshold for counting peaks: (squares) $S>1$, (triangles) $S>3$,
(circles) $S>5$.}
\label{fig:noisyMap}
\end{figure}

Using these maps as the input to our peak finding software we find that the
efficiency and completeness depend on the filter size in the expected manner.
Recalling that the $M_{\rm ap}$ filter scale is roughly $3\times$ the Gaussian
width for similarly extended kernels the results in Fig.~\ref{fig:noisyMap} can
be seen to be quite similar.
In the presence of noise the `optimal' filtering scale ($2'-4'$) is slightly
larger than for the noise free map.
One could argue that the low completeness levels we are finding are a result
of our mass threshold, and that we would find a larger fraction of the higher
mass clusters.  This is in fact true, as the middle panel in
Fig.~\ref{fig:noisyMap} shows, but the price is a very low efficiency.
In the middle panel we give the completeness and efficiency keeping only
clusters above $3\times 10^{14}h^{-1}M_\odot$
(see also Fig.~\ref{fig:s_vs_mass}).
On balance there is not a significant improvement compared to the
$10^{14}h^{-1}M_\odot$ mass threshold.
Finally we have also restricted ourselves to the clusters which lie in
regions where the lensing kernel is more than half of its peak value.  These
clusters clearly have a larger probability of contributing to a significant
peak than clusters near the observer or the sources.
The completeness and efficiency numbers are given in the lowest panel of
Fig.~\ref{fig:noisyMap}.  While the completeness is everywhere higher than
the top panel, the improvement is marginal.

\begin{figure}
\begin{center}
\resizebox{3.5in}{!}{\includegraphics{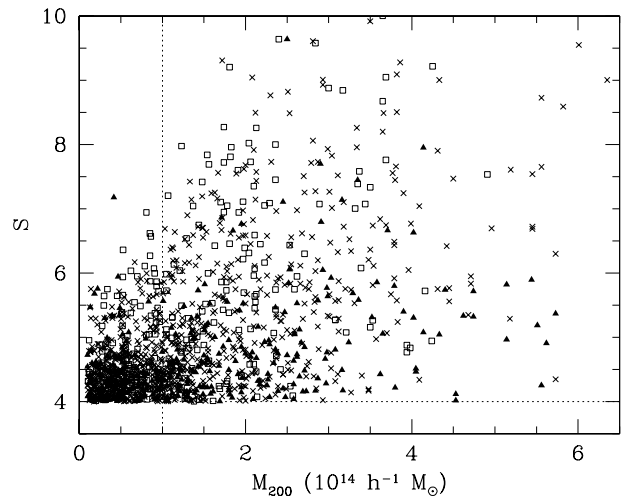}}
\end{center}
\caption{\footnotesize%
A scatter plot of unique matches of all groups with
$M_{200}>10^{13}h^{-1}M_\odot$ and all peaks with $S>4$ in our 5 maps
with a filtering scale of $\simeq 3'$.
We have divided the line-of-sight into three intervals in distance.
Open squares mark the closest $1/3$, crosses the middle $1/3$ and filled
triangles the most distant $1/3$ of the distance to the source.
Note that there are a substantial number of objects below our chosen mass
threshold ($10^{14}h^{-1}M_\odot$) for clusters.}
\label{fig:mess}
\end{figure}

Fig.~\ref{fig:noisyMap} shows that even quite high $S-$peaks can be
inefficient or incomplete.  This does not mean that these peaks are purely
noise however.  In many cases a strong peak can be found to match to a
lower mass object (or objects) along the line-of-sight.
To illustrate this we have matched all groups above $10^{13}h^{-1}M_\odot$
with all peaks having $S>4$ in the 5 fields.  Fig.~\ref{fig:mess} shows
a scatter-plot of all the unique matches.  Notice that there is a substantial
tail of lower mass groups even for $S>4$.  It is these low-mass groups that
are driving our low efficiencies.

\begin{figure}
\begin{center}
\resizebox{3.5in}{!}{\includegraphics{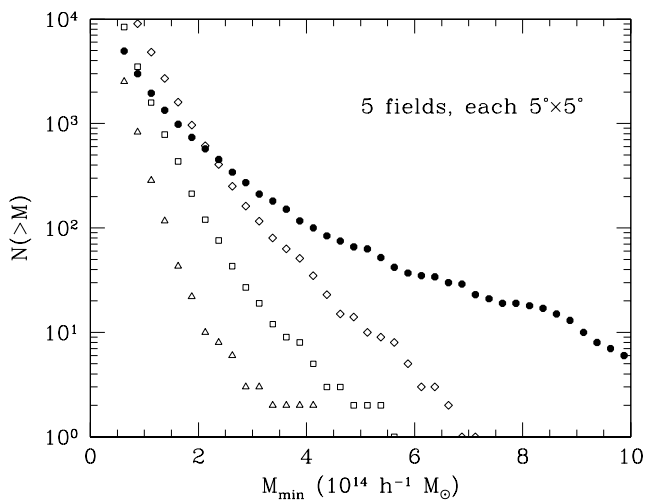}}
\end{center}
\caption{\footnotesize%
The mass functions of `real' and `missed' peaks in our noise-free maps.
Filled circles show the true mass function.
Empty symbols show the minimum estimated mass of the missed peaks (see text)
which do not match a true cluster with mass above $10^{14}h^{-1}M_\odot$.
The mass of these peaks is computed in a square box of width $2.1'$
(triangles), $2.8'$ (squares) and $3.5'$ (diamonds).}
\label{fig:miss_vs_true}
\end{figure}

Our low efficiency is caused by `noise' coming from both intrinsic
ellipticities and from cosmic structures.
The measured mass function will therefore be biased toward the most
massive objects.
This problem could be overcome if we knew the mass Probability Distribution
Function of those noise peaks, which we could try to deconvolve from the
measured mass function.
Unfortunately, only the noise peak $S$-distribution is known, as this can
be derived analytically given the ellipticity dispersion
(van Waerbeke \cite{vWae}).  
It is interesting to ask whether this problem could be suppressed with an
arbitrary low noise mass map, such as could be obtained by observing fields
with a very long exposure time.  

To answer this we measured the mass function associated with peaks
in the {\it noise-free\/} fields which do not match any real cluster above
$10^{14}h^{-1}M_\odot$.  We took peaks with positive $\kappa$ -- the mass
function would be higher if we lowered this threshold or included any noise,
it would be lower if we raised the threshold.
In order to assign a mass to a fake peak, we followed
Erben et al.~(\cite{erben00}) and computed the {\it minimum\/} mass that would
naively be associated with the peak.  This mass corresponds to the deflector
being at the redshift where the lensing kernel peaks ($z\simeq 0.5$ in our
case) and is a conservative way for observers to give a lower mass to what
they believe to be a real cluster.
Fig.~\ref{fig:miss_vs_true} shows this minimum mass function of the missed
peaks (empty symbols) compared to the true mass function (filled circles).
The estimated mass depends of course on the aperture one uses.
The three sets of empty symbols show the mass function obtained using three
different aperture sizes: we used a square of side $2.1'$, $2.8'$ and
$3.5'$, corresponding respectively to $0.7$, $1$, and $1.2~h^{-1}$Mpc radius
at the assumed lens redshift of $0.5$.
This choice matches the typical aperture size used in the literature, and
correspond to the virial radii used to compute most of the masses in
this work (Fig.~\ref{fig:clussize}).
We see that the non-matched peaks are non-negligible below
$3\times 10^{14}h^{-1}M_\odot$.
This means that even in the ideal case of noise free data, a large number
of ``clusters'' in the range [$10^{14}h^{-1}M_\odot$,
$3\times 10^{14}h^{-1}M_\odot$] are only projections of large-scale structure.
Lowering the mass threshold does not significantly change
Fig.~\ref{fig:miss_vs_true}.

Phrased another way, the presence of a distribution of halos and large-scale
structures provides a fluctuating background to our $\kappa$ maps.
The clusters we are seeking are embedded in this background, which has a
similar effect on the mass function (broadening it) as does measurement
noise.  However this broadening depends on the r.m.s.~of the mass map, which
depends on the cosmological model one considers.
Thus in order to recover the cluster mass function by deconvolution we would
need to assume an underlying cosmological model.

Also note that Fig.~\ref{fig:miss_vs_true} suggests that it is possible to
obtain peaks in a lensing map which can be interpreted at structures as
massive as a few $10^{14}h^{-1}M_\odot$ due simply to projection effects.
This may bear upon the reports of `dark clumps' with masses around a few
$10^{14}h^{-1}M_\odot$ by
(Fischer \cite{fischer99}, Erben et al.~\cite{erben00}, and
Umetsu \& Futamase \cite{UmeFut}).
In these papers, the authors computed the probability for the $S$ peak to
be `real' against random alignment of lensed galaxies, but they neglected
the effect of projection of lower mass clumps discussed here.

We show in Fig.~\ref{fig:s_vs_mass} that the efficiency and completeness
are not equally distributed among the different cluster masses.
There is a trend to have higher completeness for higher cluster masses,
as expected.  However our completeness is still not $100\%$ even for very
massive clusters.  In our 5 fields there are a handful of massive clusters with
$z\simeq 1$ and thus a low lensing efficiency.  These do not pass our $3\sigma$
threshold in Fig.~\ref{fig:s_vs_mass}.
We caution that at the high mass end we have relatively few clusters in our
maps and in the underlying simulation volume.
We thus become very sensitive to fluctuations both in the number density and
in the redshift distribution of these objects.
Thus these numbers should be taken as suggestive of some of the issues which
can affect completeness with the caveat that they should be checked with a
larger simulation before being used to `correct' any data.

We show the correlation between distance and $S$ in Fig.~\ref{fig:Dist_vs_S}.
This correlation is easy to understand: at a fixed mass low $S$ clusters
correspond preferentially to a low lensing efficiency, thus they are
at either high or at low redshift.
Since the cosmological volume is much larger at high than at low redshift,
we naturally catch more high redshift clusters.
This also points out a problem in using a fixed $S$ threshold for cluster
selection: it will affect the redshift selection function, and this should be
properly taken into account in cluster abundance analysis for instance.

\section{Comparison with previous work}

Ours is not the first study to investigate how complete a weak lensing
survey of clusters could be using numerical simulations.  It does however
improve upon early work in some respects.  The closest predecessor to our
work, in terms of focus, is that of Reblinsky \& Bartelmann (\cite{RebBar}).

These authors used N-body simulations to look at projection effects in weak
lensing selected cluster samples, and compared it to the projection effects
in richness selected clusters. They generated a 3D cluster catalog from
a single output of one of the GIF simulations in an $85 h^{-1} {\rm Mpc}$ box.
Their weak lensing maps were constructed by taking a projection of the mass
in the box, constructing the 2D lensing potential to calculate the shear, and
use aperture mass ($M_{\rm ap}$) methods to find peaks.
Since we start with a simulation volume 44 times larger (containing many more
clusters) and simulate the entire $\sim 2000h^{-1}$Mpc line-of-sight, not just
one $85h^{-1}$Mpc piece of it, we could study projection effects over
cosmological scales.
It turns out that this exacerbates the projection effects seen in their work
and makes our results slightly more pessimistic than theirs.

Reblinsky \& Bartelmann (\cite{RebBar}) worked only at $2'$.
Our efficiency/completeness at $2'$ (Fig.~\ref{fig:noisyMap}) looks similar
to their result, although it is difficult to compare the results directly
since Reblinsky \& Bartelmann (\cite{RebBar}) give only cumulative numbers.
Both sets of results can be explained by the leakage of small mass peaks to
higher masses due to the convolution of the true cluster mass function with
the `noise'.
This predicts that we should observe more high mass clusters than there
really are, in accord with our results.
Here we have demonstrated how this result depends on the smoothing kernel
used, with the dependence being non-trivial.
In particular we found that beyond a given scale (which is kernel dependent)
increasing the smoothing scale always pushes the efficiency close to $100\%$
(but the price to pay is a low completeness).

\begin{figure}
\begin{center}
\resizebox{3.5in}{!}{\includegraphics{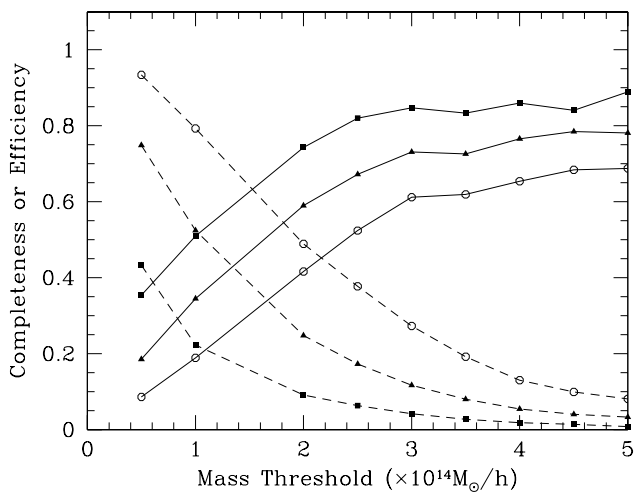}}
\end{center}
\caption{\footnotesize%
Efficiency and completeness as a function of mass for maps with a
$\simeq 3'$ filtering scale.  Solid lines show completeness, dashed lines
efficiency.  Solid squares are for $S>3$, triangles $S>4$ and open circles
$S>5$.  The statistics at the high mass end become very poor.}
\label{fig:s_vs_mass}
\end{figure}


\begin{figure}
\begin{center}
\resizebox{3.5in}{!}{\includegraphics{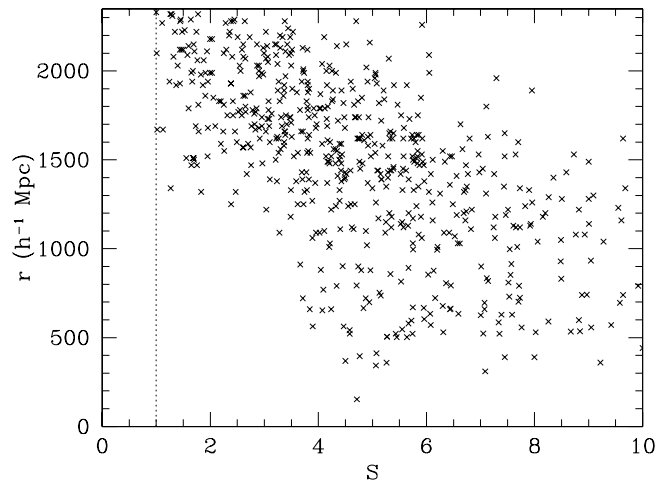}}
\end{center}
\caption{\footnotesize%
A scatter-plot showing distance and $S$ for all clusters in our 5 fields
with mass above $2\times 10^{14}h^{-1}M_\odot$ and $S>1$.
The more distant clusters, having a lower lensing efficiency, have lower
$S$ at fixed mass.
The $y$ axis runs from $z=0$ to $z=1$, with $z=0.3$ being $\sim 850h^{-1}$Mpc
{}from the observer and $z=0.5$ being $1300h^{-1}$Mpc.}
\label{fig:Dist_vs_S}
\end{figure}

Simulations which include the entire line-of-sight have been performed by
Reblinsky, Kruse, Jain \& Schneider (\cite{RKJS}), following up the
semi-analytic work of Kruse \& Schneider (\cite{KruSch}).
These authors made ``$M_{\rm ap}$ maps'' from simulated shear maps, with
and without adding random noise. From these they extracted the number
density of $M_{\rm ap}$ peaks with $S/N > 5$ (where for the map with no
noise they estimated the noise due to intrinsic galaxy ellipticities).
They find quite good agreement with the analytic estimates, in that $\sim 5$
peaks per square degree are found above this detection threshold for the map
with no noise\footnote{Note: this does {\it not\/} imply that the survey is
complete, simply that the analytic estimate produces the same fraction of
identified clusters as the simulation.}.
Unfortunately, due to the lack of knowledge of the 3D cluster positions in
their analysis, it was impossible for them to study the
completeness/efficiency.
This makes it impossible to compare directly with our work.

\section{Discussion}

In the last few years it has become possible to search for clusters of
galaxies directly as mass enhancements using weak gravitational lensing.
This method probes the mass of a cluster independent of its dynamical
state, and thus presents a different view to surveys based on galaxy
counts or the intra-cluster medium.
For this reason a lens selected survey of clusters is an appealing sample,
which could in principle be mass selected allowing a reconstruction of the
cluster mass function with redshift.  

Unfortunately there are many obstacles to be overcome.
Our study strongly implies that complementary observations (both weak and
strong lensing, optical, Sunyaev-Zel'dovich, X-ray) will be of great help in
cleaning a sample of lensing selected clusters of spurious detection and
projection effects
(e.g.~Castander \cite{Cas00}; Bartelmann \cite{Bar01}).
An example of this is the confirmation using redshifts of the lensing
selected cluster by Wittman et al.~(\cite{Witetal01}).
This will probably involve a significant number of followup observations
on lensing (mass) preselected clusters.  
In this paper we have begun to address the problem of designing such a survey,
depending on the different goals (completeness, cluster redshift, mass range)
one wants to achieve.

Extending upon the work of Reblinsky \& Bartelmann (\cite{RebBar})
and Metzler et al.~(\cite{MetWhiLok}), we studied the problems associated
with selecting clusters in lensing data, using larger numerical simulations
of clusters which simulated the entire past light-cone.
We focussed on the aperture mass statistic, assuming one identified source
population, and investigated its dependence on cluster mass and redshift.
We compared the catalogs produced by different lensing-based analyses to the
reference set of 3D clusters present in the simulation.

As discussed before, measurements of the masses of individual clusters from
weak lensing have a large scatter ($100\%$) and a significant bias
(about $20\%$), for clusters more massive than $10^{14}h^{-1}M_\odot$
(see Fig.~\ref{fig:mhist}).
The bias comes from the fact that clusters live preferentially in larger
structures.
The large scatter is due to the presence of a large number of halos of
different masses making up the large-scale structure.  Phrased in terms of
a mass, the `noise' induced by this structure is comparable to the signal from
a cluster of $10^{14}h^{-1}M_\odot$ put at redshift of $\sim 0.5$.
This may bear upon the existence of dark clusters with mass of a few
$10^{14}h^{-1}M_\odot$ in lensing surveys, however to be sure whether
projection effects are the explanation would require simulating the
distribution of the light (i.e.~galaxies) under the same observational
conditions.  If light suffers less projection than mass, this may explain
the dark clusters.
In agreement with Hoekstra (\cite{Hoe}) we find that the noise due to
{\it uncorrelated\/} large-scale structure along the line-of-sight does not
obscure the signal from sufficiently massive ($\sim 10^{15}h^{-1}M_\odot$)
clusters.

This `mass scatter' biases the mass function measured with lensing.
The scatter contains an intrinsic contribution, which is cosmological model
dependent, and a measurement noise contribution determined by the intrinsic
galaxy ellipticities and the smoothing kernel.
It is possible to model the noise in the $S$ statistic (and to predict the
number density of observed $S$ peaks as in Reblinsky et al.~(\cite{RKJS}),
for instance), but it is much more difficult to transpose this $S$-noise
into a mass noise since $S$ is not simply related to the mass
(see Fig.~\ref{fig:mess}), and this relation depends on the cosmological model. 

Even if we restrict ourselves to the methods explored here, lensing surveys
should be relatively complete for the highest mass clusters,
($\ga 3\times 10^{14}h^{-1}M_\odot$) with reasonable efficiency
($0.1-0.5$ if $S>5$ for example).
However it is important to take into account the variation of the lensing
kernel with redshift in interpreting the mass threshold of an $S$-selected
sample.

It is not clear to what extent these issues can be overcome.
In this work, we have not made use of filters matched to the cluster profiles
or of multiple source populations.  In principle incorporating either of these
could increase the efficiency or completeness of our samples.
However, at the very least these issues have to be considered in projects
aimed at performing a statistical measure of cluster masses from weak lensing.
In particular the trade-off between completeness/efficiency and mass bias are
important aspects for plans to measure the lower mass clumps, like groups of
galaxies (Schneider \& Kneib \cite{SchKne}; M\"oller et al.~\cite{MPKB}).

\section*{Acknowledgements}

We would like to thank Peter Schneider for comments on an earlier draft.
The simulations in this paper were carried out by the authors on the IBM-SP2
at the National Energy Research Scientific Computing Center.
This work was supported in part by the Alfred P. Sloan Foundation and the
National Science Foundation, through grants PHY-0096151, ACI96-19019 and
AST-9803137.


\end{document}